\documentstyle[twoside]{article}

\catcode`\@=11
\long\def\@makefntext#1{ 
\protect\noindent \hbox to 3.2pt {\hskip-.9pt
$^{{\eightrm\@thefnmark}}$\hfil}#1\hfill} 

\def\thefootnote{\fnsymbol{footnote}}
 \def\@makefnmark{\hbox to 0pt{$^{\@thefnmark}$\hss}}  

\def\ps@myheadings{\let\@mkboth\@gobbletwo
\def\@oddhead{\hbox{} 
\rightmark\hfil\eightrm\thepage}
\def\@oddfoot{}\def\@evenhead{\eightrm\thepage\hfil 
\leftmark\hbox{}}\def\@evenfoot{}
\def\sectionmark##1{}\def\subsectionmark##1{}}

\oddsidemargin=\evensidemargin
\renewcommand{\thefootnote}{\fnsymbol{footnote}}

\newcounter{sectionc}\newcounter{subsectionc}\newcounter{subsubsectionc}
\renewcommand{\section}[1] {\vspace{12pt}\addtocounter{sectionc}{1}
\setcounter{subsectionc}{0}\setcounter{subsubsectionc}{0}\noindent
	{\tenbf\thesectionc. #1}\par\vspace{5pt}}
\renewcommand{\subsection}[1] {\vspace{12pt}\addtocounter{subsectionc}{1}
	\setcounter{subsubsectionc}{0}\noindent
	{\bf\thesectionc.\thesubsectionc. {\kern1pt \bfit #1}}\par\vspace{5pt}}
\renewcommand{\subsubsection}[1]
{\vspace{12pt}\addtocounter{subsubsectionc}{1}
	\noindent{\tenrm\thesectionc.\thesubsectionc.\thesubsubsectionc.
	{\kern1pt \tenit #1}}\par\vspace{5pt}}
\newcommand{\nonumsection}[1] {\vspace{12pt}\noindent{\tenbf #1}
	\par\vspace{5pt}}

\newcounter{appendixc}
\newcounter{subappendixc}[appendixc]
\newcounter{subsubappendixc}[subappendixc]
\renewcommand{\thesubappendixc}{\Alph{appendixc}.\arabic{subappendixc}}
\renewcommand{\thesubsubappendixc}
	{\Alph{appendixc}.\arabic{subappendixc}.\arabic{subsubappendixc}}

\renewcommand{\appendix}[1] {\vspace{12pt}
        \refstepcounter{appendixc}
        \setcounter{figure}{0}
        \setcounter{table}{0}
        \setcounter{lemma}{0}
        \setcounter{theorem}{0}
        \setcounter{corollary}{0}
        \setcounter{definition}{0}
        \setcounter{equation}{0}
        \renewcommand{\thefigure}{\Alph{appendixc}.\arabic{figure}}
        \renewcommand{\thetable}{\Alph{appendixc}.\arabic{table}}
        \renewcommand{\theappendixc}{\Alph{appendixc}}
        \renewcommand{\thelemma}{\Alph{appendixc}.\arabic{lemma}}
        \renewcommand{\thetheorem}{\Alph{appendixc}.\arabic{theorem}}
        \renewcommand{\thedefinition}{\Alph{appendixc}.\arabic{definition}}
        \renewcommand{\thecorollary}{\Alph{appendixc}.\arabic{corollary}}
        \renewcommand{\theequation}{\Alph{appendixc}.\arabic{equation}}
        \noindent{\tenbf Appendix \theappendixc #1}\par\vspace{5pt}}
\newcommand{\subappendix}[1] {\vspace{12pt}
        \refstepcounter{subappendixc}
        \noindent{\bf Appendix \thesubappendixc. {\kern1pt \bfit #1}}
	\par\vspace{5pt}}
\newcommand{\subsubappendix}[1] {\vspace{12pt}
        \refstepcounter{subsubappendixc}
        \noindent{\rm Appendix \thesubsubappendixc. {\kern1pt \tenit #1}}
	\par\vspace{5pt}}

\newcommand{\textlineskip}{\baselineskip=13pt}
\newcommand{\smalllineskip}{\baselineskip=10pt}

\def\eightcirc{
\begin{picture}(0,0)
\put(4.4,1.8){\circle{6.5}}
\end{picture}}
\def\eightcopyright{\eightcirc\kern2.7pt\hbox{\eightrm c}}

\newcommand{\copyrightheading}[1]
	{\vspace*{-2.5cm}\smalllineskip{\flushleft
	{\eightrm Modern Physics Letters A, #1}\\
	{\eightrm $\eightcopyright$\, World Scientific Publishing
	 Company}\\
	 }}

\def\abstracts#1#2#3{{
	\centering{\begin{minipage}{4.5in}\baselineskip=10pt\eightrm
	\centerline{ABSTRACT}
	\parindent=0pt #1\par
	\parindent=15pt #2\par
	\parindent=15pt #3
	\end{minipage} }\par}}

\newcommand{\bibit}{\nineit}

\renewenvironment{thebibliography}[1]			
	{\ninerm
	 \baselineskip=11pt				
	 \begin{list}{\arabic{enumi}.}
	{\usecounter{enumi}\setlength{\parsep}{0pt}
	 \setlength{\leftmargin 17pt}{\rightmargin 0pt}	
	 \setlength{\itemsep}{0pt} \settowidth		
	{\labelwidth}{#1.}\sloppy}}{\end{list}}

\newcounter{itemlistc}
\newcounter{romanlistc}
\newcounter{alphlistc}
\newcounter{arabiclistc}

\newcommand{\fcaption}[1]{
        \refstepcounter{figure}
        \setbox\@tempboxa = \hbox{\eightrm Fig.~\thefigure. #1}
        \ifdim \wd\@tempboxa > 5in
           {\begin{center}
        \parbox{5in}{\eightrm \smalllineskip Fig.~\thefigure. #1 }
            \end{center}}
        \else
             {\begin{center}
             {\eightrm Fig.~\thefigure. #1}
              \end{center}}
        \fi}

\newcommand{\tcaption}[1]{
        \refstepcounter{table}
        \setbox\@tempboxa = \hbox{\eightrm Table~\thetable. #1}
        \ifdim \wd\@tempboxa > 5in
           {\begin{center}
        \parbox{5in}{\eightrm\smalllineskip Table~\thetable. #1 }
            \end{center}}
        \else
             {\begin{center}
             {\eightrm Table~\thetable. #1}
              \end{center}}
        \fi}

\def\@citex[#1]#2{\if@filesw\immediate\write\@auxout	
	{\string\citation{#2}}\fi			
\def\@citea{}\@cite{\@for\@citeb:=#2\do			
	{\@citea\def\@citea{,}\@ifundefined		
	{b@\@citeb}{{\bf ?}\@warning
	{Citation `\@citeb' on page \thepage \space undefined}}
	{\csname b@\@citeb\endcsname}}}{#1}}

\newif\if@cghi
\def\cite{\@cghitrue\@ifnextchar [{\@tempswatrue
	\@citex}{\@tempswafalse\@citex[]}}
\def\citelow{\@cghifalse\@ifnextchar [{\@tempswatrue
	\@citex}{\@tempswafalse\@citex[]}}
\def\@cite#1#2{{$\null^{#1}$\if@tempswa\typeout
	{IJCGA warning: optional citation argument
	ignored: `#2'} \fi}}

\def\pmb#1{\setbox0=\hbox{#1}
	\kern-.025em\copy0\kern-\wd0
	\kern.05em\copy0\kern-\wd0
	\kern-.025em\raise.0433em\box0}

\def\fnt#1#2{\footnotetext{\kern-.3em
	{$^{\mbox{\scriptsize #1}}$}{#2}}}

\def\fpage#1{\begingroup
\voffset=.3in
\thispagestyle{empty}\begin{table}[b]\centerline{\footnotesize #1}
	\end{table}\endgroup}

\def\runninghead#1#2{\pagestyle{myheadings}
\markboth{{\eightit{\quad #1}}\hfill}{\hfill{\eightit{#2\quad}}}}

\font\tenbf=cmbx10
\font\tenit=cmti10
\font\tenit=cmti10
\font\bfit=cmbxti10 at 10pt
 1
 1
 1

\font\ninerm=cmr9
\font\nineit=cmti9

\font\eightrm=cmr8
\font\eightit=cmti8

\def\qed{\hbox{${\vcenter{\vbox{                          
   \hrule height 0.4pt\hbox{\vrule width 0.4pt height 6pt
   \kern5pt\vrule width 0.4pt}\hrule height 0.4pt}}}$}}

\runninghead{$q$-Deformation of Quantum Spin Chains
$\ldots$} {$q$-Deformation of Quantum Spin Chains
$\ldots$}

\def\smallfrac#1#2{\mbox{\small $\frac{#1}{#2}$}}
\def\rlx{\relax\leavevmode}
\def\inbar{\vrule height1.5ex width.4pt depth0pt}
\def\IC{\rlx\hbox{\,$\inbar\kern-.3em{\rm C}$}}
\def\bm#1{\mbox{\boldmath $#1$}}
\newcommand{\emet}{{\em et al.}}

\begin{document}
\normalsize\textlineskip
{\thispagestyle{empty}
\setcounter{page}{1}

\renewcommand{\thefootnote}{\fnsymbol{footnote}} 

\copyrightheading{Vol. 0, No. 0 (1994) 000--000}

\vspace*{0.88truein}

\fpage{1}
\centerline{\bf $\bm{q}$-DEFORMATIONS OF QUANTUM SPIN CHAINS WITH}
\vspace*{0.035truein}
\centerline{\bf EXACT VALENCE-BOND GROUND STATES}
\vspace{0.37truein}
\centerline{\footnotesize M. T. BATCHELOR and C. M. YUNG}
\vspace*{0.015truein}
\centerline{\footnotesize\it Department of Mathematics,
Australian National University}
\baselineskip=10pt
\centerline{\footnotesize\it Canberra, ACT 0200, Australia}
\vspace{0.225truein}

\vspace*{0.21truein}
\abstracts{\noindent
Quantum spin chains with exact valence-bond ground states are of great
interest in condensed-matter physics. A class of such models was proposed
by Affleck \emet, each of which is $su(2)$-invariant and constructed as a sum
of
projectors onto definite total spin states at neighbouring sites. We propose to
use the machinery of the $q$-deformation of $su(2)$ to obtain generalisations
of such models, and work out explicitly the two simplest examples. In one
case we recover the known  anisotropic
spin-1 VBS model while in the other we obtain a new anisotropic generalisation
of the spin-$\smallfrac{1}{2}$ Majumdar-Ghosh model.}{}{}

\vspace*{-3pt}\textlineskip
\section{Introduction}
\noindent
The study of one-dimensional quantum spin chains
continues to flourish,
with recent applications to
diffusion-limited reactions\cite{ritt}.
In this work the integrability and quantum group
invariance of the spin chain play a crucial role,
with the quantum group deformation parameter $q$
taking on a simple physical meaning
in terms of the diffusion rates.
In this paper we pursue a different application and
take up the $q$-deformation of isotropic
quantum spin chains with exact
valence-bond ground states\cite{aklt}.

The interest in valence-bond states originates in proposed
mechanisms for high-$T_c$ superconductivity\cite{a}.
Several one-dimensional chains are known to have ground
states made up of valence-bonds. The primary examples are
the spin-$\smallfrac{1}{2}$ Majumdar-Ghosh model\cite{mg}
and the spin-1 valence-bond-solid (VBS) model of Affleck
\emet \cite{aklt} Higher-spin valence-bond models exist
on lattices with a higher co-ordination number, with e.g.,
a spin-$\smallfrac{3}{2}$ model on the honeycomb lattice.

A valence-bond between two sites on a lattice is the spin-0 (singlet)
combination of two spin-$\smallfrac{1}{2}$ states on those sites. For
spin-$\smallfrac{1}{2}$ chains it is a natural concept. In fact it makes sense
for higher spin-$s$ chains as well, since a spin-$s$ operator on one site
can be considered as the
symmetric combination of $2s$ spin-$\smallfrac{1}{2}$ operators
on the same site.
The spin-$\smallfrac{1}{2}$
Majumdar-Ghosh model\cite{mg,m} involves nearest-neighbour and
next-nearest-neighbour interactions, with Hamiltonian
\begin{eqnarray}
   {\cal H}^{\rm M-G} & = & \sum_{j=1}^{N-2} {\cal P}^{(3/2)}_{j,j+1,j+2}
     \nonumber\\
   & = & \smallfrac{2}{3}  \sum_{j=1}^{N-2} [ \bm{S}_j\cdot \bm{S}_{j+1} +
       \bm{S}_{j+1}\cdot \bm{S}_{j+2} + \bm{S}_{j}\cdot\bm{S}_{j+2} +
       \smallfrac{3}{4}],
\label{eqn:mg}
\end{eqnarray}
where ${\cal P}^{(3/2)}_{j,j+1,j+2}$ is the projector onto states with total
spin-$\smallfrac{3}{2}$ at sites $j$, $j+1$ and $j+2$.
The ground states of this model are dimerised,
as depicted in Figure 1 for a lattice with $N=7$ sites, with each
dimer representing a valence-bond. That these are indeed ground states, of
zero energy, is most evident from the $su(2)$-projector description of the
Hamiltonian, since if there is a valence-bond shared between
every three neighbouring
sites the total spin on those sites can only be $\smallfrac{1}{2}$. For a
lattice with even (odd) number of sites, the ground state
degeneracy is five (respectively, four) in the case of free boundary
conditions, although there are two different ground states in the infinite
lattice limit.

\vskip 0.5cm
\begin{center}
\setlength{\unitlength}{0.005250in}%
\begingroup\makeatletter
\def\x#1#2#3#4#5#6#7\relax{\def\x{#1#2#3#4#5#6}}%
\expandafter\x\fmtname xxxxxx\relax \def\y{splain}%
\ifx\x\y   
\gdef\SetFigFont#1#2#3{%
  \ifnum #1<17\tiny\else \ifnum #1<20\small\else
  \ifnum #1<24\normalsize\else \ifnum #1<29\large\else
  \ifnum #1<34\Large\else \ifnum #1<41\LARGE\else
     \huge\fi\fi\fi\fi\fi\fi
  \csname #3\endcsname}%
\else
\gdef\SetFigFont#1#2#3{\begingroup
  \count@#1\relax \ifnum 25<\count@\count@25\fi
  \def\x{\endgroup\@setsize\SetFigFont{#2pt}}%
  \expandafter\x
    \csname \romannumeral\the\count@ pt\expandafter\endcsname
    \csname @\romannumeral\the\count@ pt\endcsname
  \csname #3\endcsname}%
\fi
\endgroup
\begin{picture}(486,106)(74,637)
\thicklines
\put( 77,740){\circle*{8}}
\put(157,740){\circle*{8}}
\put(237,740){\circle*{8}}
\put(317,740){\circle*{8}}
\put(397,740){\circle*{8}}
\put(477,740){\circle*{8}}
\put(557,740){\circle*{8}}
\put(157,640){\circle*{8}}
\put(397,640){\circle*{8}}
\put(477,640){\circle*{8}}
\put(557,640){\circle*{8}}
\put( 77,640){\circle*{8}}
\put(237,640){\circle*{8}}
\put(317,640){\circle*{8}}
\put(77,740){\line( 1, 0){ 80}}
\put(237,740){\line( 1, 0){ 80}}
\put(397,740){\line( 1, 0){ 80}}
\put(157,640){\line( 1, 0){ 80}}
\put(315,640){\line( 1, 0){ 80}}
\put(477,640){\line( 1, 0){ 80}}
\end{picture}
\end{center}
\begin{center}
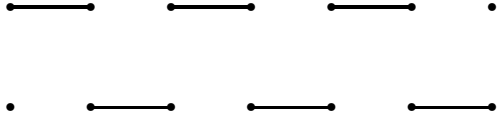
{\small Fig. 1. Dimerised ground states of the Majumdar-Ghosh model}
\end{center}
\vskip .2cm

\noindent
On the other hand, the spin-1 VBS model involves only nearest-neighbour
interactions and has the Hamiltonian
\begin{eqnarray}
   {\cal H}^{\rm VBS} & = & \sum_{j=1}^{N-1} {\cal P}^{(2)}_{j,j+1}
     \nonumber\\
   & = & \smallfrac{1}{6}\sum_{j=1}^{N-1} [ (\bm{S}_{j}\cdot \bm{S}_{j+1})^2
       + 3\bm{S}_j\cdot \bm{S}_{j+1} +
       2],
\label{eqn:vbs}
\end{eqnarray}
where ${\cal P}^{(2)}_{j,j+1}$ projects onto states with total spin $2$
at sites $j$ and $j+1$.  In this case the
groundstates are such that the valence-bonds cover the whole lattice,
as depicted in Figure 2. Once again, the nature of the ground states is clear
from the projector description of the Hamiltonian, since if there is a
valence-bond between any two neighbouring sites, the four
spin-$\smallfrac{1}{2}$'s at those sites can only add to spin 0 or 1. For this
model, the ground state degeneracy is four for finite lattices (with free
boundary conditions) with a unique ground state in the infinite lattice limit.

\vskip 0.5cm
\setlength{\unitlength}{0.005250in}%
\begin{center}
\begin{picture}(486,6)(74,637)
\thicklines
\put(160,640){\circle*{8}}
\put(400,640){\circle*{8}}
\put(480,640){\circle*{8}}
\put(560,640){\circle*{8}}
\put( 80,640){\circle*{8}}
\put(240,640){\circle*{8}}
\put(320,640){\circle*{8}}
\put(160,640){\line( 1, 0){ 80}}
\put(320,640){\line( 1, 0){ 80}}
\put(480,640){\line( 1, 0){ 80}}
\put( 80,640){\line( 1, 0){ 80}}
\put(240,640){\line( 1, 0){ 80}}
\put(400,640){\line( 1, 0){ 80}}
\end{picture}
\end{center}
\begin{center}
{\small Fig. 2. Valence-bond ground state of the VBS model}
\end{center}
\vskip .2cm

An anisotropic version of the VBS model was recently proposed
in Ref.\ 6 as a special case of the most general
$U_q(su(2))$-invariant spin-1 quantum chain.
This $q$-deformed version of the VBS model was subsequently
investigated by Kl\"umper \emet \cite{kaz1}
In this paper we present a natural derivation of this more
general model in the framework of $U_q(su(2))$-projectors.
We also use this machinery to derive a new generalisation of the
spin-$\smallfrac{1}{2}$ Majumdar-Ghosh model.
Similar results can be obtained for the higher spin VBS models of
Affleck \emet \cite{aklt,r}
However, the spin-$\smallfrac{1}{2}$ and spin-1 cases are special in
that the $U_q(su(2))$ spin matrices are proportional to their $su(2)$
counterparts and thus the $q$-deformed versions of these models can be
viewed as anisotropic generalisations.

The paper is arranged as follows. In section 2.1 we rederive the
well-known $U_q(su(2))$-invariant spin-$\smallfrac{1}{2}$ XXZ chain.
This is the simplest possible $U_q(su(2))$-invariant chain, and although
not possessing  valence-bond ground states,
its construction illustrates well the general procedure we follow
and serves as
a warm up to the derivation of the $q$-deformed
Majumdar-Ghosh model in section 2.2, which constitutes our main result.
Then in section 3 we give the derivation of the $q$-deformed spin-1 VBS model.

To first set the notation, we
review some relevant facts concerning the quantum algebra
$U_q(su(2))$.\cite{kr,d,j}
This algebra is generated by $S^z$ and $S^{\pm}$, with
defining relations $[S^+,S^-]=[2S^z]$ and $[S^z,S^{\pm}]=\pm S^{\pm}$,
where $[x]\equiv (q^x-q^{-x})/(q-q^{-1})$. The co-product $\Delta :
U_q(su(2)) \rightarrow U_q(su(2))\otimes U_q(su(2))$ given by
\begin{eqnarray}
\Delta(S^z) & = & 1\otimes S^z + S^z\otimes 1\\
\Delta(S^{\pm}) & = & q^{-S^z}\otimes S^{\pm} + S^{\pm}\otimes q^{S^z}
\end{eqnarray}
is a homomorphism : $\Delta(ab)=\Delta(a)\Delta(b)$ and is co-associative :
$(1\otimes \Delta)\circ
\Delta = (\Delta\otimes 1)\circ \Delta$.
It has a natural extension to $\Delta^{(n)} : U_q(su(2))\rightarrow
U_q(su(2))^{\otimes (n+1)}$, defined recursively
by $\Delta^{(n)}=((1^{\otimes (n-1)})\otimes\Delta)\circ \Delta^{(n-1)}$ and
$\Delta^{(1)}=\Delta$.

The Casimir element, belonging to the centre of $U_q(su(2))$,
is given by $C=S^-S^+ +[S^z+\smallfrac{1}{2}]^2 -
[\smallfrac{1}{2}]^2$, which in the limit $q\rightarrow 1$ becomes
the familiar $\bm{S}\cdot\bm{S}=(S^x)^2+(S^y)^2+(S^z)^2$,
where $S^x, S^y$ are such that $S^{\pm}=S^x \pm i S^y$. When $q$ is not a
root of unity, the representation theory of $U_q(su(2))$ is exactly the
same as that of $su(2)$. Namely, the irreducible representations are labelled
by the spin $j$ and are $(2j+1)$-dimensional. On the spin-$j$ representation
the Casimir $C$ takes the value $[j][j+1]$. This equivalence between the
representation theory of the $q$-deformed and undeformed algebras makes it
possible to $q$-deform an $su(2)$ model of Affleck \emet \cite{aklt}
with exact valence-bond
ground states by simply replacing $su(2)$-projectors with their $q$-analogues.
These projectors can be obtained from the Casimir $C$; more specifically,
 as polynomials in
$\Delta^{(n)}(C)$ where $n$ is determined by the range of interactions
required -- $n=1$ for nearest-neighbour interactions, $n=2$ for nearest-
and next-nearest-neighbour interactions, etc.

\section{Spin-half chains}

\subsection{The XXZ model}
\noindent
Let $V=\IC^2$ be a spin-$\smallfrac{1}{2}$ module for $U_q(su(2))$. The
generators
$S^z$ and $S^{\pm}$ in this representation take the same form as
their (undeformed) $su(2)$ counterparts, namely
\begin{displaymath}
S^z=\left(\begin{array}{cc}\smallfrac{1}{2} & 0\\0 & -\smallfrac{1}{2}
  \end{array}\right),\hspace{10pt}
S^+=\left(\begin{array}{cc}0 & 1\\0 & 0\end{array}\right),\hspace{10pt}
S^-=\left(\begin{array}{cc}0 & 0\\1 & 0\end{array}\right).
\end{displaymath}
{}From the definition of the co-product $\Delta$ and the Casimir $C$
we can calculate the element
$C^{(2)}\equiv\Delta(C)\in U_q(su(2))^{\otimes 2}$. On the tensor product
of two spin-$\smallfrac{1}{2}$ representations, it takes the form
\begin{displaymath}
C^{(2)}=\left(\begin{array}{cccc}q+q^{-1}&0&0&0\\
  0& q^{-1} & 1 & 0\\
  0& 1 & q & 0\\
  0&0&0& q+q^{-1}\end{array}\right).
\end{displaymath}
This can be expressed alternatively as
\begin{eqnarray}
C^{(2)}&=&2(S^x\otimes S^x+S^y\otimes S^y) + \left(q+q^{-1}\right)
S^z\otimes S^z
  +\nonumber\\
& & \smallfrac{1}{2}\left(q^{-1}-q\right)(S^z\otimes 1 - 1\otimes S^z)
  +
  \smallfrac{3}{4}\left(q+q^{-1}\right) (1\otimes 1)
\end{eqnarray}
in terms of the spin matrices.
Let $h^{\rm XXZ}\equiv C^{(2)} - \smallfrac{3}{4}\left(q+q^{-1}
  \right)1\otimes 1$ and consider the object
\begin{equation}
{\cal H}^{\rm XXZ} = \sum_{j=1}^{N-1}1\otimes\cdots\otimes 1\otimes
  h_{j,j+1}^{\rm XXZ}\otimes 1 \otimes \cdots \otimes 1,
\end{equation}
where $h_{j,j+1}^{\rm XXZ}$ is a copy of $h^{\rm XXZ}$ acting
in the $(j,j+1)$ slot of $V^{\otimes N}$. It turns out that
${\cal H}^{\rm XXZ}$ is a special case
of a more general (integrable) Hamiltonian for the
XXZ quantum chain with open boundaries and surface fields,
and can be written in more usual notation as\cite{abbbq,ps}
\begin{equation}
{\cal H}^{\rm XXZ} = \sum_{j=1}^{N-1}2(S^x_jS^x_{j+1}+S^y_jS^y_{j+1})
  + \left(q+q^{-1}\right)S^z_jS^z_{j+1} +
  \smallfrac{1}{2}\left(q^{-1}-q\right) (S_1^z-S^z_N),
\end{equation}
where $S^z_j$,  $S^x_j$ and $S^y_j$
are copies of $S^z$, $S^x$ and $S^{y}$ acting in
the $j$-th slot of $V^{\otimes N}$.
By construction, it is $U_q(su(2))$-invariant, i.e.
\begin{displaymath}
[{\cal H}^{\rm XXZ},{\cal S}^z]=[{\cal H}^{\rm XXZ},{\cal S}^{\pm}]=0,
\end{displaymath}
where
\begin{eqnarray}
  {\cal S}^z &=& \Delta^{(N-1)}(S^z) = \sum_{j=1}^{N}
1\otimes\cdots\otimes 1\otimes
  S^z_j\otimes 1 \otimes \cdots \otimes 1,\nonumber\\
{\cal S}^{\pm} &=& \Delta^{(N-1)}(S^{\pm}) = \sum_{j=1}^{N}
q^{S^z}\otimes\cdots\otimes q^{S^z}\otimes
  S^{\pm}_j\otimes q^{-S^z} \otimes \cdots \otimes q^{-S^z},
\end{eqnarray}
generate $U_q(su(2))$ on $V^{\otimes N}$. Since on $V\otimes V$ the
centre of $U_q(su(2))$ is spanned by $C^{(2)}$ and $1\otimes 1$, if we
replace $h^{\rm XXZ}$ by either of the projectors ${\cal P}^{(0)}$ or
${\cal P}^{(1)}$ (in line with the other constructions of spin chains
in this paper)
we would end up with an equivalent (up to trivial additive and multiplicative
constants) quantum chain.
It is a happy accident that this most general
nearest-neighbour spin-$\smallfrac{1}{2}$
quantum chain with $U_q(su(2))$ symmetry is integrable.

\subsection{The $q$-deformed Majumdar-Ghosh model}
\noindent
As described in the Introduction, the Majumdar-Ghosh model (\ref{eqn:mg})
can be constructed as a sum of $su(2)$-projectors
${\cal P}^{(3/2)}_{j,j+1,j+2}$. Its
$q$-deformation is evidently given by
\begin{equation}
{\cal H}^{\rm qM-G} = \sum_{j=1}^{N-2} {\cal P}^{(3/2)}_{j,j+1,j+2},
\end{equation}
where now ${\cal P}^{(3/2)}$ is a projector onto the
spin-$\smallfrac{3}{2}$ subspace
in the decomposition of the $U_q(su(2))$-module $V\otimes V\otimes V$.
As noted also in the Introduction,
the equivalence of the representation theory for $U_q(su(2))$
and $su(2)$ (when $q$ is not a root a unity)
implies that  all the arguments for
the fact that the Majumdar-Ghosh model has exact valence-bond ground states
remain valid for the $q$-deformation.

To obtain the projector ${\cal P}^{(3/2)}$ we first construct the
matrix representative of $C^{(3)}\equiv
\Delta^{(2)}(C)$, which has eigenvalues $[\smallfrac{1}{2}][\smallfrac{3}{2}]$
and $[\smallfrac{3}{2}][\smallfrac{5}{2}]$,
each with multiplicity four, being eigenvalues on irreducible
spin-$\smallfrac{1}{2}$ (two copies) and spin-$\smallfrac{3}{2}$
modules respectively. It is then clear that the projector is
\begin{equation}
{\cal P}^{(3/2)} = \frac{C^{(3)} - [\smallfrac{1}{2}][\smallfrac{3}{2}]
    (1\otimes 1 \otimes 1)}{[\smallfrac{3}{2}][\smallfrac{5}{2}]-
    [\smallfrac{1}{2}][\smallfrac{3}{2}]}.
\end{equation}
The element $C^{(3)}$ in $U_q(su(2))^{\otimes 3}$ can be constructed
in an analogous way to $C^{(2)}$, using the
definitions of $C$ and $\Delta$.  In the spin-$\smallfrac{1}{2}$
representation the result
can be written in the form
\begin{eqnarray}
  C^{(3)} & = & \left(q+q^{-1}\right)(1\otimes S^x\otimes S^x +
    1\otimes S^y\otimes S^y + S^x\otimes S^x \otimes 1 + S^y\otimes S^y
    \otimes 1) \nonumber\\
  & + & 2(q-q^{-1})(S^x\otimes S^x\otimes S^z + S^y\otimes S^y\otimes S^z -
      S^z\otimes S^x\otimes S^x - S^z\otimes S^y\otimes S^y)\nonumber\\
  & + & 2( S^x \otimes 1 \otimes S^x +S^y \otimes 1 \otimes S^y +
    S^z \otimes 1 \otimes S^z)\nonumber\\
  & + & \left(q^2+q^{-2}\right)( 1\otimes S^z\otimes S^z+
    S^z\otimes S^z \otimes 1) \nonumber\\
  & + & \smallfrac{1}{2}\left( q^2 - q^{-2}
    \right)(1\otimes 1\otimes S^z - S^z \otimes 1\otimes 1) \nonumber\\
  & + & \frac{(1+q^2)(1+q+q^2)^2}{2q^2(1+q)^2}(1\otimes 1\otimes 1).
\end{eqnarray}
It then follows that the Hamiltonian for the $q$-deformed Majumdar-Ghosh model
is
\begin{eqnarray}
{\cal H}^{\rm qM-G} &=& \frac{1}{1+q^2+q^{-2}}\left[\sum_{j=1}^{N-2}\left\{
    \left(q+q^{-1}\right)(\bm{S}_j\cdot\bm{S}_{j+1} +
     \bm{S}_{j+1}\cdot\bm{S}_{j+2}) + 2 \bm{S}_j\cdot\bm{S}_{j+2}
  \right. \right.\nonumber\\
  &+&  2\left(q-q^{-1}\right)
     \left( (S^x_jS^x_{j+1}+S^y_jS^y_{j+1})S^z_{j+2}) -
S^z_j(S^x_{j+1}S^x_{j+2}
     + S^y_{j+1}S^y_{j+2})\right)\nonumber\\
  &+& \left.\left(q^2+q^{-2}-q-q^{-1}\right)
     (S^z_{j+1}S^z_{j+2} + S^z_jS^z_{j+1})\right\} \nonumber\\
  &+&  \smallfrac{1}{2}\left(q^2-q^{-2}\right)(S^z_N+
      S^z_{N-1}-S^z_1-S^z_2) {\mbox{\Huge $]$}} + \smallfrac{1}{2}(N-2).
\label{eqn:qmg}
\end{eqnarray}
Note the presence of boundary terms, as in the XXZ chain, and also of
three-spin interactions. The constant term can of course be dropped, but is
naturally present to make the ground state energy zero for chains of all
length $N$. This generalisation of the Majumdar-Ghosh model differs from
that given by Shastry and Sutherland which includes only two-spin
interactions\cite{ss}, and which is presumably not $U_q(su(2))$-invariant.

The $q$-deformed Majumdar-Ghosh Hamiltonian is $U_q(su(2))$-invariant by
construction. We note that the most general
spin-$\smallfrac{1}{2}$
$U_q(su(2))$-invariant Hamiltonian with next-nearest neighbour interactions
can be constructed
by starting from  a linear combination of $C^{(3)}$ and $(C^{(3)})^2$ (which
together with the identity generate the centre of $U_q(su(2))$ on
$V^{\otimes 3}$). Amongst all such chains one can expect to find one which is
integrable, whose periodic version is the next highest conserved quantity
of the XXZ Hamiltonian.

\section{The $q$-deformed spin-1 VBS model}
Now let $V=\IC^3$, a spin-1 module for $U_q(su(2))$. In this representation,
$S^z$ and $S^{\pm}$ take the form
\begin{displaymath}
S^z=\left(\begin{array}{ccc}1 & 0 & 0 \\0&0&0\\0 & 0 &-1
  \end{array}\right),\hspace{10pt}
S^+=\sqrt{[2]}\left(\begin{array}{ccc}0 & 1 &0\\0 & 0& 1\\0&0&0
    \end{array}\right),\hspace{10pt}
S^-=\sqrt{[2]}\left(\begin{array}{ccc}0 & 0&0\\1 &
0&0\\0&1&0\end{array}\right).
\end{displaymath}
As described in the Introduction, the spin-1
VBS model (\ref{eqn:vbs}) of Affleck \emet \cite{aklt}
can be constructed as a sum of $su(2)$
projectors ${\cal P}^{(2)}$.
Once again, its $q$-deformation is clear; being
\begin{equation}
{\cal H}^{\rm qVBS} = \sum_{j=1}^{N-1} {\cal P}^{(2)}_{j,j+1},
\end{equation}
where ${\cal P}^{(2)}$ is now a projector onto the spin-2 subspace in the
decomposition of $V\otimes V$ as a $U_q(su(2))$-module.

As before, ${\cal P}^{(2)}$ can be obtained from the Casimir $C$, this
time as
\begin{equation}
{\cal P}^{(2)} = \frac{C^{(2)}}{[2][3]}
  \left(\frac{C^{(2)}-[2]1\otimes 1}{[2][3]-[2]}\right),
\end{equation}
since $0$, $[2]$ and $[2][3]$ are the eigenvalues of $C^{(2)}\equiv \Delta(C)$
on irreducible modules of spin 0,1 and 2, respectively. The construction of
$C^{(2)}$ goes as before, via the definitions of $\Delta$ and $C$.
Its matrix representative in the present case is a lot more
laborious to work out than in the spin-$\smallfrac{1}{2}$ case,
with the result expressible as
\begin{eqnarray}
C^{(2)} &=& (q+q^{-1})(S^x\otimes S^x + S^y\otimes S^y)
   -\smallfrac{1}{2}(q+q^{-1})^2(q-q^{-1})
   (S^z\otimes 1 -1 \otimes S^z)\nonumber\\
&+& \smallfrac{1}{4}(q+q^{-1})^3 S^z\otimes S^z +
   \smallfrac{1}{2}(q+q^{-1})(q-q^{-1})^2\left((S^z)^2\otimes 1 +
   1\otimes (S^z)^2\right)\nonumber\\
& + & (2-q-q^{-1})\left\{(1\otimes S^z)(S^x\otimes S^x + S^y\otimes S^y)
   (S^z\otimes 1) -\right.\nonumber\\
&  & \left.(S^z\otimes 1)(S^x\otimes S^x + S^y\otimes S^y)
   (1\otimes S^z)\right\}\nonumber\\
&  - &
    \smallfrac{1}{2}(q-q^{-1})\left\{(S^x\otimes S^x + S^y\otimes S^y)
    (S^z\otimes 1 -1 \otimes S^z)-\right.\nonumber\\
& & \left.(S^z\otimes 1 -1 \otimes S^z)
    (S^x\otimes S^x + S^y\otimes S^y) \right\}\nonumber\\
&  - & \smallfrac{1}{4}(q+q^{-1})^2(q-q^{-1})\left((S^z)^2\otimes S^z -
    S^z\otimes (S^z)^2\right)
   + 2(q+q^{-1})(1\otimes 1),
\end{eqnarray}
in terms of the spin matrices.
If we let $C^{(2)}_{j,j+1}=(q+q^{-1})(S^x_jS^x_{j+1}+S^y_jS^y_{j+1}) + \cdots$
be a copy of $C^{(2)}$ acting on the $(j,j+1)$ slot of $V^{\otimes N}$ then the
Hamiltonian for the $q$-deformed VBS model is given by
\begin{equation}
{\cal H}^{\rm qVBS}=\left([2]^2[3]([3]-1)\right)^{-1}\sum_{j=1}^{N-1}
   C^{(2)}_{j,j+1}\left(C^{(2)}_{j,j+1}-[2]\right).
\label{eqn:qvbs}
\end{equation}
Since the matrix representatives of $S^z$, $S^{x}$ and $S^{y}$ in the spin-1
representation  are
proportional to their (undeformed) $su(2)$ counterparts, the Hamiltonian
${\cal H}^{\rm qVBS}$
can be written wholly in terms of the latter after appropriate rescalings.

The Hamiltonian (\ref{eqn:qvbs})
(or rather, one equivalent to it up to an additive and a
multiplicative constant) was first proposed in Ref.\ 6
as potentially having exact valence-bond ground states. In that paper, the
most general $U_q(su(2))$-invariant spin-1 Hamiltonian was written down as
\begin{displaymath}
 {\cal H}^{\rm spin-1}= \sum_{j=1}^{N-1}{\cal O}_{j,j+1}(a,b),
\end{displaymath}
with the Hamiltonian on two sites ${\cal O}(a,b)$ being the most general
linear combination of
$C^{(2)}$ and $(C^{(2)})^2$ (which together with the identity generate the
centre of $U_q(su(2))$ in $V\otimes V$). The operator ${\cal O}(a,b)$ has three
eigenvalues $E_k(a,b)$ corresponding to the three spin-$k$
representations ($k=0,1,2$) in the decomposition of $V\otimes V$. The
Hamiltonian proposed corresponds to a choice of parameters $a=\tilde{a}$
and $b=\tilde{b}$
such that $E_0(\tilde{a},\tilde{b})=E_1(\tilde{a},\tilde{b})$.
Since one can always shift ${\cal O}(\tilde{a},\tilde{b})$
by a constant to make $E_0(\tilde{a},\tilde{b})=E_1(\tilde{a},\tilde{b})=0$,
it is essentially the projector ${\cal P}^{(2)}$. The
projector nature of this special choice of ${\cal O}(a,b)$ was first
discussed in Ref.\ 7 where the ground state properties of
${\cal H}^{\rm qVBS}$ were calculated. In particular, the ground state
is unique, there is a finite gap to excitations, and correlations decay
exponentially. We turn now to the properties of
the $q$-deformed Majumdar-Ghosh model.

\section{Properties of the $q$-deformed Majumdar-Ghosh model}
\noindent
The ground state correlation functions of the Majumdar-Ghosh
model are particularly simple due to the exact dimerised
ground states.\cite{m,aklt,th} For ease of comparison with the
earlier results, we consider the generalised Hamiltonian (\ref{eqn:qmg})
for even $N$ with periodic boundary conditions, so that the surface terms
ensuring the $U_q(su(2))$ invariance vanish. In this case the ground
state is two-fold degenerate with eigenvectors
\begin{eqnarray}
\phi_1 &=& \left[1,2\right] \left[3,4\right] \cdots \left[N-1,N\right],
\\
\phi_2 &=& \left[2,3\right] \left[4,5\right] \cdots \left[N,1\right],
\end{eqnarray}
where
\begin{equation}
\left[l, m\right] = \frac{1}{\sqrt{1+q^2}} \left(\uparrow_l \downarrow_m
- \, q \downarrow_l \uparrow_m\right).
\end{equation}
For each state the ground state energy is zero. The ground state
wave function is given by either of the functions
$\Psi^{\pm} = \phi_1 \pm \phi_2$,
which ensure translational symmetry. The matrix element
$\langle \phi_1 | \phi_2 \rangle$ vanishes as
\begin{equation}
\langle \phi_1 | \phi_2 \rangle = (-1)^{N/2} \,
2 \left( q + q^{-1} \right)^{-N/2},
\end{equation}
so the states $\phi_1, \phi_2$ become orthogonal in the thermodynamic
limit for $q$ real. On the other hand,
$\langle \phi_1 | \phi_1 \rangle = \langle \phi_2 | \phi_2 \rangle =1$.

Various spin-spin correlations can be considered. In particular, the
correlation associated with N\'eel order is defined by
\begin{equation}
C_n^\pm = \langle \Psi^\pm | S_1^z S_{1+n}^z | \Psi^\pm \rangle.
\end{equation}
In the $N \rightarrow \infty$ limit with $q=1$,
$C_1^\pm=-\smallfrac{1}{2}$ with $C_n^\pm = 0$ for $n > 1$,
indicating the absence of N\`eel long-range order.\cite{m,th}
However, in the same limit away from $q=1$,
$C_n^\pm = (-1)^n \smallfrac{1}{2}$, independent of $q$.
Thus the anisotropy induces a form of long-range N\`eel order.

As a final remark, we note that the model remains massive as a function of
$q$. In general there is a symmetry in the eigenspectrum about the value
$q=1$ with the eigenvalues satisfying $E(q) = E(1/q)$. Thus confining our
attention to the region $0 \le q \le 1$, we see that the gap $\Lambda$
opens up from the value\cite{th,st}
$\Lambda \simeq \smallfrac{2}{3}(0.236)$ at $q=1$
to the exact value $\Lambda=2$ (even $N$) or $\Lambda=1$ (odd $N$)
at $q=0$ where the Hamiltonian is trivial.

\nonumsection{Acknowledgements}
\noindent
This work has been supported by the Australian Research Council.

\nonumsection{References}

\end{document}